\documentclass[12pt,preprint]{aastex}
\usepackage{emulateapj5,graphicx,times}

\slugcomment{Submitted to AJ}

\shorttitle{Extended ionized gas clouds in z=0.4 cluster}
\shortauthors{Yagi et al.}
\begin{document}

\title{Discovery of nine extended ionized gas clouds in a z=0.4 cluster}

\author{Masafumi Yagi\altaffilmark{1,2},
Liyi Gu\altaffilmark{3,4},
Yusei Koyama\altaffilmark{2,5},
Fumiaki Nakata\altaffilmark{6},
Tadayuki Kodama\altaffilmark{2},
Takashi Hattori\altaffilmark{6},
Michitoshi Yoshida\altaffilmark{7}
}

\altaffiltext{1}{email:YAGI.Masafumi@nao.ac.jp}
\altaffiltext{2}{
Optical and Infrared Astronomy Division,
National Astronomical Observatory of Japan,
2-21-1, Osawa, Mitaka, Tokyo, 181-8588, Japan}
\altaffiltext{3}{
Reserch Center for Early Universe, School of Science,
The University of Tokyo, 7-3-1 Hongo, Bunkyo-ku, Tokyo 113-0011, Japan}
\altaffiltext{4}{Department of Physics, The University of Tokyo, 
7-3-1 Hongo, Bunkyo-ku, Tokyo 113-0011, Japan}
\altaffiltext{5}{Institute of Space and Astronautical Science,
Japan Aerospace Exploration Agency, 3-1-1 Yoshinodai,
Chuo-ku, Sagamihara, Kanagawa, 252-5210, Japan}
\altaffiltext{6}{Subaru Telescope, 650 North A'ohoku Place, Hilo, 
Hawaii 96720, USA}
\altaffiltext{7}{
Hiroshima Astrophysical Science Center, Hiroshima University,
1-3-1, Kagamiyama, Higashi-Hiroshima, Hiroshima, 739-8526, Japan
}

\def\Ha{H$\alpha$}
\def\Hb{H$\beta$}
\def\Hg{H$\gamma$}
\def\Hd{H$\delta$}
\def\He{H$\epsilon$}

\def\HI{\ion{H}{1}}
\def\HII{\ion{H}{2}}
\def\OII{[\ion{O}{2}]}
\def\OIII{[\ion{O}{3}]}
\def\NII{[\ion{N}{2}]}
\def\SII{[\ion{S}{2}]}
\def\HeI{[\ion{He}{1}]}

\def\RAform#1#2#3#4{#1{$^{\rm h}$}#2{$^{\rm m}$}#3{$^{\rm s}$}#4}
\def\Decform#1#2#3{#1$^{\circ}$#2'#3''}

\begin{abstract}
From deep \Ha\ imaging data of Suprime-Cam/Subaru, 
we discovered nine 
extended ionized gas clouds 
(EIG) around galaxies in Abell 851 cluster (A851) at z=0.4. 
We surveyed 30 $\times$ 25 arcmin region,
and the EIGs
were found only near the cluster center 
($<$2.3 arcmin $\sim$ 750 kpc).
The parent galaxies of the EIGs are star-forming or post-starburst
galaxies, all of which are spectroscopically confirmed members of
the cluster.
Four out of the nine parent galaxies show distortion
of stellar distribution in the disk, 
which can be a sign of recent
interaction, and the interaction may have made EIGs. 
On the other hand, six parent galaxies (one overlaps those exhibiting distortion)
show \Ha\ emission without stars, which implies a ram pressure stripping.The spectrum of the brightest parent galaxy 
shows a post-starburst signature, and resembles the \Ha\ stripped 
galaxies found in the Coma cluster.
Meanwhile, two brightest parent galaxies in A851 are more massive than
the EIG parent galaxies in the Coma cluster. 
This is consistent with ``downsizing'' of star-forming galaxies, 
though it is still in a statistical fluctuation.
We also analyzed Suprime-Cam data of another z=0.39 cluster, CL0024+17, 
but found no EIGs. 
The key difference between A851 and CL0024+17 would be
the existence of a subcluster colliding with the main body of A851,
in which six or seven out of the nine parent galaxies in A851 exist,
and the fraction of EIGs in the subcluster is significantly higher than
the main subcluster of A851 and CL0024+17.
\end{abstract}


\section{Introduction}

A deep \Ha\ imaging gives us a chance
to detect an extended ionized gas clouds (EIGs) out of galaxies.
EIGs were found in several nearby (z$<$0.1) clusters
\citep[e.g.,][]{Gavazzi2001,Yoshida2002,Cortese2006,Yagi2007,
Sun2007,Kenney2008,Yagi2010,Sun2010,ArrigoniBattaia2012,Fossati2012,
Kenney2014}.
The gas would have been stripped from a galaxy (parent galaxy)
and observed in \Ha\ emission.
Though the mechanism of stripping 
and ionizing is not yet fully understood,
it is thought that 
a galaxy interaction \citep{Toomre1972}
and/or
ram-pressure stripping \citep[RPS;][]{Gunn1972}
would have made EIGs in some cases.

The progenitor of EIG parents would be gas-rich and 
therefore probably star-forming galaxy before the stripping.
The stripping would weaken or cease the star formation
and the galaxy would transition into post-star-formation and 
eventually quiescent phase.
It is widely known that the fraction of 
blue galaxies, which are thought to be gas-rich and star forming galaxies,
in a cluster increases with the redshift at z$>$0.2
\citep[Butcher-Oemler effect;][]{Butcher1984}.
This implies that the major transition from a star-forming galaxy
to a quiescent galaxy 
occurred in cluster at z$>$0.2.
Moreover, many post-starburst galaxies 
are found in z$\sim$0.4 clusters \citep{Poggianti1999}
compared with local clusters.
Though gas stripping is not the only mechanism for the transition,
we can expect that gas-stripping events
would have been more frequent in distant clusters of galaxies.
Nevertheless, no distant EIGs have been discovered so far.
It is not clear whether such EIGs are quite rare at z$\sim$0.4,
or simply because no searches have been carried out yet.

In this paper, we report the detection of 
nine
EIGs in Abell 851(A851), 
and compare their parent galaxies with those found in the Coma cluster.
We adopted WMAP 9-year cosmology 
($h_0$,$\Omega_M$,$\Omega_\lambda$)=(0.697,0.282,0.718)
\citep{Hinshaw2013}.
The magnitudes are given in AB-system \citep{Oke1983}.
The redshift is given in the local standard of rest frame.

\section{Imaging Data}

\subsection{Target Clusters}
We searched for extended \Ha\ emitters at z$\sim$0.4 in 
Suprime-Cam\citep{Miyazaki2002} data
in the public archive of the Subaru Telescope (SMOKA)%
\footnote{\url{http://smoka.nao.ac.jp/}}\citep{Baba2002}.
Two clusters were observed both in broadband (W-S-Z+; z-band) 
and in narrowband corresponding to \Ha:
A851 at z=0.405 in N-B-L921(NB921) 
and CL0024+17 at z=0.390 in N-B-L912(NB912).
The redshifts of the clusters are taken from
\citet{Oemler2009} and \citet{Moran2007}.
The distance modulus is 
m-M=41.74(A851) and m-M=41.64(CL0024+17).
The angular scales of 5.47 kpc arcsec$^{-1}$(A851)
and 5.33 kpc arcsec$^{-1}$(CL0024+17) are adopted.
The effective band transmission for z, NB912, and NB921 are
(center,FWHM)=(9043\AA,948\AA), 
(9137\AA,130\AA), and (9194\AA,130\AA), respectively.
The redshift coverage for \Ha\ emission is therefore 
z=0.392$\pm$0.010(NB912) and 0.401$\pm$0.010(NB921), respectively.
\citet{Hayashino2003} reported the
non-uniformity of NB921's transmission function.
This effect is investigated in Appendix B. 
Objects with redder continuum tend to 
have z-NB921 or z-NB912 color larger and may mimic the emission,
because the center wavelength of NB921 and NB912 is redder than that of z-band.
We therefore used R-band (W-C-RC; center=6500\AA,FWHM=1170\AA) data
to check the continuum color.

\subsection{Reduction of Imaging Data}

For the study of star-forming galaxies,
the z and NB921 data of A851 were once analyzed by \citet{Koyama2011},
and z and NB912 data of CL0024+17 were
analyzed by \citet{Kodama2004}.
We re-analyzed the data with a correction of fringe pattern,
which was skipped in the previous analyses.
The detail of the fringe correction is given in Appendix A.
The flux zero-point of z-band was calibrated using SDSS3 DR9 catalog
\citep{DR9}, as the method by \citet{Yagi2013}.
The Galactic extinction at A851 is A$_{\rm V}$=0.046 from NED,
based on \citet{Schlafly2011}.
We adopted A$_{z}$=0.020, and used it for NB921, too.
The 1-$\sigma$ limiting surface brightness in 2 arcsec aperture 
was 27.3 mag arcsec$^{-2}$ for both z and NB921.
The flux measurement results of z-band and NB921 were
consistent with \citet{Koyama2011}.
In the following analysis, we used the photometry
of B, R and z band by \citet{Koyama2011},
and z-NB921 from the new analysis.
For the search of EIGs in A851,
we first estimated a model color of galaxies at z=0.405 as z-NB921$\sim$0.05,
using the theoretical spectral energy distribution(SED)
of passive galaxies from \citet{Furusawa2000},
which is based on \citeauthor{Kodama1997}'s \citeyearpar{Kodama1997} model.
We therefore constructed z-band subtracted NB921 image 
(NB-z image, hereafter), 
so that the region where z-NB921=0.05 is to be gray. 
The slight difference in the point spread function 
makes the center of the bright object positive. 
However, it does not affect the detection of EIGs.

For CL0024+17, we estimated z-NB912$\sim$0.03 at z=0.390.
CL0024+17 was processed in the same way as A851 and 
z-band subtracted NB912 image (we also call it the NB-z image, hereafter)
was constructed, so that z-NB912=0.03 is gray.
We adopted A$_{z}$ and A$_{NB912}$=0.070 as the Galactic extinction.

A drizzled image of A851 center taken in F814W with 
the {\it Advanced Camera for Surveys} of the {\it Hubble Space Telescope}
(ACS/HST)
was retrieved from the Mikulski Archive for Space Telescopes 
(MAST archive\footnote{\url{http://archive.stsci.edu/}}).
F814W filter covers 7000--9500\AA,
which corresponds to 5000--6800\AA\ in the rest frame.
The data were used to see the morphology of the parent galaxies
in A851.

\subsection{Detection of EIG candidates}

Extended \Ha\ features were searched by an eye-inspection in the NB-z image,
and checked in a three color composite image of
R(blue), z(green), and NB921(red).
We identified nine candidates of EIGs.
The list of their parent galaxies is sorted by the z-band magnitude
(Table \ref{tab:target}).
It should be stressed that we surveyed 30 $\times$ 25 arcmin region,
and the EIG candidates were found only around the cluster center 
($<$2.3 arcmin).
The cutouts from the NB-z image, the three color composite image,
and the ACS image in F814W are shown as Figure \ref{fig:postage}.
In left panels, the NB-z excess is indicated by a red contour.
The surface brightness of the contour corresponds to 3 $\sigma$ 
of the NB-z image in 2 arcsec aperture,
which is NB921 26.2 mag arcsec$^{-2}$
if no continuum is contaminated.
The green contour in the panels shows the isophote,
which encloses 90-\% of the total flux in z-band.
The EIG candidates from eye-inspection showed an extension of the 
NB-z excess (red contour) over the 90-\% isophote.
In the right panels, F814W band images are shown.
Thanks to the high spatial resolution of ACS, 
we can see that most of the EIGs' parent galaxies 
have a clumpy morphology.

As the extended NB-z excess features 
can be an accidental overlap of a distant
object, we cannot conclude they are EIGs at z=0.4 without 
a spectroscopy of them.
In the EIG study in the Coma cluster, however, all the 14 
candidates were found to be actual EIGs in the Coma cluster
(M. Yoshida et al. in preparation).
It implies that the extended narrowband excess features
without continuum are highly likely to be EIGs.

In our analysis, EIG candidates were only found in A851, 
and no extended \Ha\ emission was found in CL0024+17.
We therefore focused on A851 in the next section.

\section{Properties of the EIGs and the parent galaxies in A851}

\subsection{Follow-up Spectroscopy}

Five out of the nine parent galaxies (C1--C5)
are known spectroscopic members of A851 
confirmed by \citet{Oemler2009} and/or \citet{Nakata2014},
while the others (C6, C7, C8 and C9) 
did not have spectroscopic redshifts.
We therefore performed a spectroscopic observation 
on 2013 October 31 UT with 
the Faint Object Camera and Spectrograph 
\citep[FOCAS;][]{Kashikawa2002} at the Subaru Telescope.
We used multi-object spectroscopy mode, 
300R grism and 
SY47 order-cutting filter, which cover the wavelength range 
approximately from 4700\AA\ to 10000\AA. 
The slit widths were 0.8 arcsec
resulting in the spectral resolution of $\sim$ 10\AA.
We obtained four 10-minutes exposures.
The wavelength was calibrated using 
night sky lines.
In C6, C7, and C9, we identified emission lines, 
and the redshifts were determined by them.
C8 did not show emission, but strong absorption features 
of \Hg, \Hd, CaH+\He, and CaK, which enabled us to
determine the redshift.
The measured redshifts are given in Table \ref{tab:target}.
C6, C7, C8 and C9 had a redshift $\sim 0.395$.
In summary, the nine parents of the EIG candidates were 
confirmed to be a member of A851 clusters.
We therefore assume that the nine EIG candidates are actually 
EIGs in A851 hereafter,
though the final confirmation of 
the physical association between the parents and the EIGs 
requires further spectroscopy of EIGs, 
as tried for Coma EIGs in \citet{Yagi2007} and \citet{Yoshida2012}.

We also obtained spectra of C1 and a bright spot of C2
(in the middle panel of Figure \ref{fig:postage}, 
 a magenta clump seen in the north-northeast of the parent galaxy).
The redshift of C1 was measured from \OII, \Hb, \Ha, 
and \NII\ emissions, and \Hd\ absorption to be z=0.4065$\pm$0.003.
The value is comparable to previous studies; 
z=0.4064 \citep{Nakata2014}, and z=0.4060 \citep{Oemler2009}.
The redshift of the C2 bright spot showed 
\OII, \Hb, \OIII, \Ha, and \NII.
The measured redshift of the C2 bright spot was z=0.4063
which is comparable to the redshift of the parent galaxy;
z=0.4061 by \citet{Oemler2009}.

\subsection{Nature of EIG parents}
\label{sec:nature}

In Figure \ref{fig:CMD}, color-magnitude diagram(CMD) 
of member galaxies selected by photometric redshift (phot-z) 
in \citet{Koyama2013} are plotted in z vs R-z plane.
The color of parent galaxies is given in Table \ref{tab:color}.
The z vs R-z at z$\sim$0.4 corresponds to R vs B-R 
in the rest-frame 
at z$\sim$0, 
and is comparable to the R vs B-R plot and the r vs g-r plot 
of nearby clusters (e.g., Fig. 8 of \citealp{Yagi2010}) 
except for the zero-point offsets.

In Figure \ref{fig:CMD}, the distribution of the EIG parents shows
an offset from that of the other \Ha\ emitters (red filled circles);
EIG parents tend to have bluer R-z color and/or brighter z magnitude.
Apparently the colors of EIG parents
do not show a correlation with z-band magnitude,
which was observed in the Coma cluster \citep{Yagi2010}.
In Figure \ref{fig:CMD_NB}, narrowband color excess (z-NB) 
vs z-band is plotted. 
Some parents show strong color excess while C1 and C8 show 
no excess.
In FOCAS spectra, C1 and C8 show strong Balmer absorption,
and resemble post-starburst (so-called E+A or k+a) galaxy.
The variety of the color and the narrowband excess 
suggests that EIG parents in A851 are a mixture of 
different origin and/or different stage of the stripping.

Our new spectra of C6--C9, and the bright spot of C2 
showed \HII-region-like line ratios.
The \NII/\Ha\ vs \OIII/\Hb\ plot is shown as 
Figure \ref{fig:BPT} and the data are given in Table \ref{tab:lineratio}.
The measurement method of the line ratios is
the same as \citet{Yagi2013}.
C7 shows strong Balmer absorptions of 
stars of the galaxy.
Using GALAXIV stellar synthesis model \citep{Bruzual2003},
we estimated that equivalent widths 
of the star absorption would be no larger than 14\AA\ and 8\AA\ at
\Ha\ and \Hb, respectively.
The correction of this extreme case is shown as 
a vector in Figure \ref{fig:BPT}.
Even after the correction, C7 has
\HII-region-like line ratios.
Four of the NB-z excess parents (C2--C4,C7)
has data in GALEX object catalog, and the UV-optical color 
in the observed frame is blue (NUV-R=0.2--1.5).
It also suggests that the stellar component is young.
Moreover, the bright six galaxies (C1--C6) show Spitzer 24$\mu$m emission
larger than 200 $\mu$Jy 
in the wide-field data used in \citet{Koyama2013}.
For a check of a possible active galactic nuclei (AGN) contribution
to the 24$\mu$m emission,
we used the Spitzer IRAC flux in
Spitzer Enhanced Imaging Products Source List.
C1--C7 have flux data in 4 bands (Table \ref{tab:color}),
and their two-color diagram is shown in Figure \ref{fig:IRcolmag}.
Their colors were out of the empirical region of AGN
by \citet{Stern2005} except for a known AGN (C2).
Though C1 is near the AGN region, it is not an AGN,
since the FOCAS spectrum of C1 does not show a sign of AGN;
it shows a very weak \OIII\ emission, and an 
\Hb\ emission in a broad \Hb\ absorption.
The 24$\mu$m emission would therefore be
yet another sign of recent star formation in C1 and C3--C6.
In the ACS data, almost all the parent galaxies have disky morphology;
some of them have highly disturbed spirals and the others look less
disturbed(Figure \ref{fig:postage}). 

The stellar mass (M$_{*}$)of the parents was estimated using the recipe
in \citet{Koyama2013}.
They verified that this method works reasonably well
for star-forming galaxies (with an uncertainty of $\sim$ 0.2 dex level) by
comparing with the value derived from a multi-band SED fitting.
We adopted the equations for z=0.4;\begin{equation}
\log(M_{*}/10^{11}M_{\odot})=-0.4(z-20.07)+\Delta \log M,
\end{equation}
\begin{equation}
\Delta \log M=0.054-3.81 \exp[-1.28 (B-z)].
\end{equation}
The estimated mass is given in Table \ref{tab:target}.
It should be noted that C2 has AGN and the mass estimation
has larger uncertainty than others.
The EIG parent galaxies in A851 are 
relatively more massive than those in the Coma cluster \citep{Yagi2010}.
The stellar mass is 10$^{9.4}$ -- 10$^{11.8}$ M$_{\odot}$, and 
they would be giant galaxies.

In summary, most of the EIG parents in A851 are star-forming
spiral galaxies, and the others are post-starburst galaxies.
The EIGs found in A851 are therefore expected to be 
a stripped or ejected gas from 
the parent star-forming/post-starburst galaxy.

\subsection{Distribution of parent galaxies}
\label{sec:parentdist}

The spatial distribution of the parents is shown in Figure \ref{fig:map}.
The EIGs were somehow detected only in 
the northern side of the cluster.
The field of view of the ACS data is also 
shown in Figure \ref{fig:map}.
Note that the eye inspection was done in NB-z and R, z, NB images, 
and the ACS image that mainly covered the northern part did not affect
the north-south inhomogeneity of EIGs.

\citet{Oemler2009} discussed that
 A851 consists of several subclusters.
In Figure \ref{fig:zd}, the projected distance from the cluster center
versus relative recession velocity to the cluster (z=0.405) is plotted.
The different symbol represents member of each subcluster.
\citet{Oemler2009}
classified C1 and C2 to be the member of 
the cluster main body (``Core'' subcluster),
and C3 and C5 to be the member of northern subcluster (``North'' subcluster).
Figure \ref{fig:zd} shows that most of the EIG parents 
except for C1 and C2 overlap the North subcluster members
(open circles). 
C6 has a comparable redshift to the North subcluster 
while a Core member also exists near C6.
In Figure \ref{fig:map}, the celestial position of C6 is near 
the center of the cluster, while other North subcluster members
exist in the northern part.
We therefore assume that most of the parent galaxies unclassified 
by \citet{Oemler2009} are members of the North subcluster,
while C6 belongs to either the Core subcluster or the North subcluster.

\subsection{Individual Parent Galaxies}

In this section, we investigate each EIG parent.
We checked two features that would provide
some hints on the origin of each EIG by visual inspection.
One of the hints is a distortion of the stellar distribution in 
the disk.
Since RPS affects gas but not the stars,
it basically does not move stars,
while galaxy interaction will make apparent tidal features.
However, it is not a crucial criterion, since there is a simulation
that shows RPS can change the distribution of stars
\citep{Vollmer2003,Steinhauser2012}.
\citet{Vollmer2003} showed that
the movement of gas changes the gravitational field, 
and stars are drifted.
In the simulation by \citet{Steinhauser2012},
stars formed in the tail fall back to the parent galaxy,
go through the disk, and change the stellar distribution.
The other hint is whether \Ha\ emission without stars exists.
Since tidal force affects both stars and gas,
EIG would overlap with the stripped stars.
An example is \Ha\ of Stephan's quintet (HCG92).
On the other hand, RPS works only to the gas in principle,
and sometimes makes a long \Ha\ tail without stars.
As new stars formed in the stripped gas 
are sometimes found in EIGs
\citep[e.g.,][]{Yoshida2008,Yagi2013b},
the overlap of stars does not always mean interaction,
but \Ha\ emission without stars, 
especially distant from the parent galaxy 
suggests RPS.
More detailed observational data are required to investigate further.
Diagnostic line ratios and a kinematic structure of EIGs 
are useful parameters to reveal the nature of EIGs
\citep[e.g.,][]{Yoshida2008},
which will be obtained by future spectroscopic observation of EIGs.
In this work, we just showed the two features.

\subsubsection{C1}
The EIG from C1 was a long ($>$80 kpc) tail without 
a stellar counterpart.
Moreover no obvious distortion of the disk is recognized. 
The tail would have been made by RPS.

As mentioned in Section \ref{sec:nature},
C1 was classified as a post-starburst (k+a) galaxy \citep{Oemler2009}.
\citet{Nakata2014} measured the equivalent width of \OII\ of C1
as EW(\OII)=11.3\AA; they classified it as e(a), emission with 
strong Balmer absorption. 
Our follow-up spectroscopy revealed that 
the core of C1 (central 1 arcsec aperture)
has a strong (the equivalent width $\sim$ 100\AA) \Ha+\NII\ emission,
\Hb\ emission in a broad absorption, and no detectable \OIII\ emission.
The difference implies that star formation in C1 is
compact and slit spectrum largely affected by slit position and aperture.
C1 would be a galaxy with a central star-forming region 
in a post-starburst disk.
Such coexistence was reported in some post-starburst galaxies
\citep[e.g.,][]{Matsubayashi2011,Swinbank2012}.
C1 shows an emission in Spitzer 24$\mu$m data.
\citet{Dressler2009} argued that such an IR detected k+a 
had a strong burst 2--5 $\times 10^8$ yr before the observation
epoch and has ongoing star formation.
The galaxy would be in the final phase of a starburst to change 
into a post-starburst phase.

In the ACS image in F814W, there was a sign of narrow streams
toward the same direction as the \Ha\ tail 
(Figure \ref{fig:C1}),
but their positions do not always overlap the \Ha\ emission.
The streams may be a stellar component that was made in situ,
like the ``fireball'' features found in the Coma 
cluster\citep{Yoshida2008,Yoshida2012}.
However no compact \Ha\ emission was recognized in the tail.
The difference can be partly explained by the low S/N and 
low resolution of A851 data as discussed in Section \ref{sec:massComa}.

Our NB-z image shows elongated \Ha\ from the center and 
possible \Ha\ absorption in the disk, which is consistent with the 
spectroscopic classification.
Such post-starburst galaxies with elongated \Ha\ tail were also found 
in the Coma cluster, such as GMP2910,
GMP3071, GMP3779 \citep{Yagi2007,Yagi2010}.
The stellar mass of C1 ($6\times 10^{11}$ M$_{\odot}$) is,
however, much larger than those in Coma ($10^9$--$10^{10}$M$_{\odot}$).

\subsubsection{C2}
The EIG from C2 includes a bright spot near the galaxy,
and extended and clumpy \Ha\ emission to the northeast.
The line ratios from our FOCAS spectrum showed that
the bright spot would be an \HII\ region (Figure \ref{fig:BPT}).

In the ACS image in F814W shows a sign of stars 
along the \Ha\ extension.
This resembles the fireball features found in the Coma 
cluster\citep{Yoshida2008,Yoshida2012}.
The spectra of C2 is classified as e(n) by \citet{Oemler2009}
following the classification by \citet{Dressler1999}.
and consistent with the result from IRAC/Spitzer color 
discussed in the previous section.
Strong 24$\mu$m emission of C2 (3.5 mJy) 
would be because of the AGN.
Since the EIG has a spatial offset from the core, 
and the spectrum of the spot in the EIG shows 
signs of \HII\ region, this EIG is unlikely to be of AGN origin.
C2 shows an asymmetry of stellar 
distribution.
On the other hand, it has extended \Ha\ emission that seems
to have no association with stellar emission.
\subsubsection{C3}
C3 shows loose and disturbed spiral arms, 
and the \Ha\ emitting regions along the arms.
Its appearance implies that it is interacting with 
a neighbor spiral at northeast.
The projected distance between them is 50 kpc.
EIG may have been created by the interaction,
though the redshift of the neighbor is currently not available.
C3 has a redshift of z=0.3956,
and classified as a member of North subcluster by \citet{Oemler2009}.

\subsubsection{C4}
C4 shows a flocculent disk, and smooth extension of 
\Ha\ toward the southeast.
EIG may be a result of interaction with C5.
The difference of the recession velocity of C4 and C5 
is somehow as large as 720 km s$^{-1}$.
Though C4 is not cataloged in \citet{Oemler2009},
its position and redshift
suggest that it is a member of North subcluster.

\subsubsection{C5}
In the ACS image, C5 shows several remote and compact objects
in the south-southwest, which are thought to be star clusters.
The direction is the same as the \Ha\ extension. 
The distortion of the disk, the remote star clusters and 
the \Ha\ emission may be a result of the interaction with C4.
However, it shows a slight sign of \Ha\ emission without stars 
in the southeast direction.C5 is cataloged as a member of the North subcluster.
The spectrum of C5 is classified as e(a) by \citet{Oemler2009}.
The spectral class is thought to be a dust-obscured starburst
\citep{Poggianti1999,Dressler1999,Dressler2009,Oemler2009}.

\subsubsection{C6}
EIG is barely recognized toward the west. 
There is no counterpart in F814W image, 
and therefore it could be a pure gas-stripping event by RPS.
FOCAS spectrum showed strong Balmer emissions, (\Ha, \Hb, \Hg\ and 
\Hd\ emissions), and \OII, \SII\ and a sign of \HeI\ 5876 emissions.
As shown in Figure \ref{fig:zd}, 
C6 could be a member of either the North subcluster or 
the Core subcluster.

\subsubsection{C7}
Three remote \Ha\ regions are recognized at southeast.
Its appearance resembles that of some dwarf galaxies in the Coma cluster
(e.g., GMP4232) and those found in z=0.2 clusters \citep{Cortese2007}.
The FOCAS spectrum of the parent galaxy showed \Hd\ and \Hg\ emissions 
in absorption, indicating that it is in a transition from
a star-forming phase to a quiescent phase.
As discussed in Section \ref{sec:nature},
the strong Balmer absorption would also affect 
\OIII/\Hb\ and \NII/\Ha, 
but the diagnostic line ratios after maximum correction
are still comparable to those of \HII\ regions.
Since the ACS/HST image did not cover C7,
it is unclear whether the stellar distribution in the disk 
shows a distortion or not.\subsubsection{C8}
In the ACS image, a chain of two galaxies is recognized
to the south of C8. \Ha\ emission follows the chain, and then 
turns to the west, where no stellar counterpart is seen.
In the FOCAS spectrum of C8, 
the north galaxy shows the typical signatures of a post-starburst galaxy;
blue continuum, strong \Hd\ absorption, and no emissions.
\subsubsection{C9}
C9 is a compact spiral with a warped disk in the ACS image.
The \Ha\ extension is toward the northeast without a stellar counterpart, 
and resembles ram-pressure stripped objects.
The parent galaxy showed \Ha, \Hb, and \Hg\ emissions.
In FOCAS spectrum, this spatial extension was also recognized in 
\Ha\ and \Hb.

\section{Discussion} 

\subsection{Properties of parent galaxies: comparison with Coma}
\label{sec:massComa}

In A851, massive ($> 10^{11}$ M$_{\odot}$) 
EIG parents are found (C1 and C2), 
which were not found in the Coma \citep{Yagi2010}.
Meanwhile, less massive and faint parents ($< 10^{9}$ M$_{\odot}$) 
are not found.

The lack of faint parents in A851 is explained by the detection limit.
The 1-$\sigma$ limiting surface brightness in 2 arcsec aperture 
of the Coma data was 27.5 mag arcsec$^{-2}$
in N-A-L671(NA671) \citep{Yagi2010},
and (center,FWHM) of NA671 are (6712\AA,120\AA) \citep{Yagi2007}.
If we consider the cosmological dimming and the difference of 
the filter widths, it corresponds to 28.4 mag arcsec$^{-2}$ 
for \Ha\ emission in NB921.
Therefore, A851 data, whose limiting surface brightness 
is 27.3 mag arcsec$^{-2}$ in NB921, is 1.1 mag shallower
for the same \Ha\ emitting object in the Coma.
We binned and blurred the Coma data of \citet{Yagi2010}
so that the limiting surface brightness and resolution 
are comparable to the A851 data quality and checked the detectability.
Only three or four (GMP2559, GMP3816, GMP3896 and marginally
GMP2910) out of fourteen 
will be recognized as extended \Ha\ objects.
The estimated mass of the four Coma EIG parents are
$\sim 10^{10}$ M$_{\odot}$, and three show star formation 
in the disk.
These features resemble the EIG parents found in A851.
The absence of detection of $< 10^9$ M$_{\odot}$ EIG parents is therefore
explained by the shallower limiting magnitude of the A851 data.

The lack of bright EIG parents in the Coma is, meanwhile,
not explained by the difference of the data quality. 
One possibility is that the typical EIG parents 
have changed since z=0.4. 
The progenitor of an EIG parent 
would be an infalling galaxy from the surrounding field.
At z=0.4, a larger number of massive star-forming galaxies 
existed around clusters\citep[e.g.,][]{Kodama2001}.
At z$\sim$0, many of such massive star-forming galaxies 
around clusters
have already fallen into the cluster, 
and probably have lost their gas.
Moreover, the number of
massive star-forming galaxies has decreased even in low-density fields,
because of the cosmic evolution.
Thus massive EIG parent would be rare at z$\sim$0.
\citet{Poggianti2004} argued that the post-starburst galaxies 
in the Coma cluster is less massive than those in z=0.4 clusters.
This ``downsizing effect'' is consistent with the result 
that a massive and post-starburst EIG parent exists in A851.
Another possible answer is that it is simply a statistical fluctuation.
If the expected number of bright EIG parents in a cluster
is two, for example, the probability of no detection is $\sim$ 14\%.
EIG search in more clusters will enable us to distinguish the two
possible hypotheses.

Regarding the spatial distribution,
the detected EIGs are located at a projected distance of 0.2--0.8 Mpc
from the center of the cluster (Figure \ref{fig:zd}).
The range is comparable to that of Coma EIGs 
\citep{Yagi2010}.
\citet{Smith2010} surveyed 
gaseous stripping events in the Coma cluster using UV images,
and they found that all except the least certain case (GMP5422)
exist in 0.2--0.9 Mpc from the center.
It suggests that the gaseous stripping also occurs 
in comparable environments.
The results imply that EIG would have 
strong correlation with a cluster environment,
even though the origin of the stripping seems to be in interaction.
Galaxy interaction may assist an effective gas stripping 
in infalling galaxies, as it would destabilize the gas in the galaxy.

\subsection{Lack of EIGs in CL0024+17}
In previous sections, we investigated EIGs in A851.
In CL0024+17, we found no EIGs.
There were several extended features in the NB912-z image, 
but they were all in the disk of galaxies.
CL0024+17 is said to consist of several subclusters
\citep{Czoske2002}. 
Two subclusters (A and B) are colliding
head-on along the line of sight. The subclusters have redshift of
0.395(A) and 0.381(B), and our narrowband NB912 covered the 
\Ha\ at the redshift of both subclusters.
The non-detection was therefore not because of 
the wider redshift distribution.

We check the image qualities of CL0024+17 and A851.
The 3 sigma detection in NB912-z for CL0024+17
corresponds to 26.2 mag arcsec$^{-2}$ in NB912,
which is comparable to the detection in A851;
26.2 mag arcsec$^{-2}$ in NB921. 
The seeing size of the CL0024+17 data ($\sim$ 1.2 arcsec)
was better than that of A851 data ($\sim$ 1.5 arcsec).
Therefore, no detection of EIGs in CL0024+17 
would not be artificial, and it really lacks EIGs.

We then examine whether the difference of A851 and CL0024+17 
could be explained by a statistical fluctuation.
It is natural to assume that the number of EIG parents 
follows Poisson statistics, and the expected number $\lambda$ 
would be proportional to the number of the member galaxies 
in the cluster.

From statistics,
N=0 can be reproduced with p-value$>0.05$ if $0<\lambda<3.0$,
where $\lambda$ is the expected value of the Poisson distribution.
Meanwhile, N=9 sets $5.5<\lambda<13.8$.
Therefore, the $\lambda$, the expected number of EIG parents,
of A851 must be larger than that of CL0024+17, at least by 
a factor of 1.8.

The number ratio of the member galaxies in the two clusters are
calculated by statistical background subtraction
\citep[e.g.,][]{Binggeli1988}.
The projected number density of $-24<M_z<-19$ objects 
at projected distance $>$ 1.5 Mpc from the center of the cluster was 
used as the background distribution.
The density was subtracted from the projected number 
density of $\leq$1.5 Mpc.
As CL0024+17 is thought to be a cluster merger along the line of sight
\citep{Czoske2002}, the mass estimation from X-ray properties 
or velocity dispersion may be affected by the configuration
\citep[e.g.,][]{Umetsu2010}, and not suitable for comparison with A851.
Our simple number count would therefore be a good method to estimate 
the cluster richness, because it is even robust against 
the complicated geometry, and because the Poisson statistics are additive.

The background corrected projected number densities are
57$\pm$5 Mpc$^{-2}$ and 46$\pm$5 Mpc$^{-2}$ 
for A851 and CL0024+17, respectively.
The number ratio of the cluster N(A851)/N(CL0024+14) is 
estimated as 1.2$\pm$0.2.
In the analysis, we adopted the center coordinate as 
\RAform{09}{42}{57.46},\Decform{+46}{58}{49.8} for A851
and  
\RAform{00}{26}{35.67},\Decform{+17}{09}{43.1} 
for CL0024+17 \citep{Wen2013}. 
We also tried different coordinates given in NED,
but the result was not affected.

It should also be noted that there are still many \Ha-emitting 
galaxies in the two clusters \citep{Kodama2004,Koyama2011}.
\citet{Koyama2011} argued that the total star-formation rate within 
0.5$\times$ $R_{200}$ is comparable in the two clusters.
As $R_{200}$ of CL0024+17 \citep[1.7 Mpc;][]{Kodama2004} 
is smaller than that of A851 \citep[2.13 Mpc;][]{Koyama2011},
more star-forming galaxies should exist in CL0024+17 within a fixed radius.
The lack of EIGs in CL0024+17 is therefore not due to the lack of 
gas rich parent galaxies.

We can therefore conclude that the difference of the 
EIG fraction of A851 and CL0024+17 is significant.
It cannot be explained by the statistical fluctuation.

\subsection{EIG Enhancement in Infalling Subcluster}
In A851, 2(or 3, depending on the membership of C6) 
of the parents would be a member of the Core subcluster,
and the other 7(or 6) would be that of the 
North subclusters.
Meanwhile, the richness of the North subcluster is supposed
to be $\sim$ 1/8 of the Core subcluster, 
as the number ratio of confirmed members in \citet{Oemler2009} is 11/80.
The recession velocity dispersions of the subclusters are 
1079 km s$^{-1}$ (Core) and 295 km s$^{-1}$, respectively
\citep{Oemler2009}.
These results suggest that the North subcluster would be
smaller than the Core subcluster by an order of magnitude.

We examined the significance of the difference of 
N=3(Core) and N=6(North) in the number ratio of 1/8. 
In this test, we assume 
that C6 is a member of the Core subcluster, 
since the null hypothesis of the test is 
that the difference is not significant.
N=3 can be reproduced with p-value$>0.05$ if $1.0<\lambda<7.0$,
and N=6 can be reproduced with p-value$>0.05$ if 
$3.0<\lambda<10.5$.
If $\lambda$ is proportional to the number of members of each 
subcluster, the difference is significant.
The EIG fraction is therefore quite high in the North subcluster.
The difference of A851 Core subcluster and CL0024+17 is,
meanwhile, within the statistical fluctuation,
with $1<\lambda<3$. 

The high EIG fraction rate in an infalling group 
resembles the case in Abell 1367 at z=0.02, where 
many EIGs were detected around Blue Infalling Group\citep{Cortese2006}.
In the Coma cluster, the correlation between
EIG parents and infalling group was not recognized, but 
EIG parents tend to have larger deviation in recession velocity 
\citep{Yagi2010}.
The relative line of sight velocity of the group with respect to the cluster
core is as large as $\sim$ 3000 km s$^{-1}$. 
During the course of its violent infall/merger onto the Core subcluster,
some star-forming galaxies in the North subcluster may have been 
or are experiencing some environmental effects such 
as RPS and/or tidal interactions with neighbors, 
which then produced the signatures of EIGs.

\subsection{Age of C1 tail}

The spectrum of C1 shows a post-starburst signature. 
If the termination of the starburst was caused by 
the stripping, it occurred within $\sim 10^9$ years.

If we assume that the beginning of the stripping 
is seen as the tip of \Ha\ tail, we can date the age.
Since peak-to-peak recession velocity is $\sim$ 2$\times$2000 km s$^{-1}$ 
(Figure \ref{fig:zd})
and the relative recession velocity of C1 to the Core subcluster 
is $\sim$ -100 km s$^{-1}$, the tangential velocity of C1 
is assumed to be $\sim \sqrt{2000^2-100^2} \sim 2000$ km s$^{-1}$.
The projected length of the EIG, 86 kpc, 
corresponds to movement for 4$\times 10^7$ years with the speed,
and the far tip of the tail would be stripped from 
the parent 4$\times 10^7$ years ago in that case.
Though the age is barely enough to
cease the \Ha\ emission \citep[e.g.,][]{Yagi2013b},
the \Ha\ tail near the parent should have been stripped more recently,
and it would be difficult to stop \Ha\ emission widely in the disk.
The spatially extended Balmer absorption 
requires a much longer time, e.g., $\sim 2\times 10^8$ yr.

One possible solution is that
the tangential velocity is much smaller than 2000 km s$^{-1}$.
If the tangential velocity is $\sim$ 400 km s$^{-1}$, for example,
the tail length corresponds to $\sim 2\times 10^8$ yr.
Another possibility is that the stripped gas is more extended 
than the apparent \Ha\ tail. 
Such extended gas clouds were found in \HI\ or in X-ray 
\citep[e.g.,][]{Oosterloo2005,Sun2007,Gu2013}.
The comparison of the length of \Ha\ tail and 
in other wavelengths is important for understanding the 
ionizing source of the \Ha\ emission and the fate of the
stripped gas.

\section{Summary}

We have discovered nine EIGs in a z=0.4 cluster (A851).
All their parents are spectroscopically 
confirmed members of the cluster,
and they all lie near the cluster center 
($<$2.3 arcmin $\sim$ 750 kpc).

Six or seven of the parent galaxies would be members of the
northern group of galaxies,
the North subcluster \citep{Oemler2009},
which is probably just merging onto the 
core of the cluster main body (Core subcluster).
In another z=0.4 cluster, CL0024+17, we found no EIG candidates.
The relative EIG fraction is significantly high in the North 
subcluster, while that of the A851 Core subcluster and CL0024+17
are comparable within a statistical fluctuation.
The emergence of EIGs in the North subcluster of A851 
may thus be related to the infall of
the subcluster, and the gas stripping therein due to 
some environmental effects.

Two massive ($> 10^{11}$ M$_{\odot}$) 
parent galaxies of EIGs are found in A851,
The fact that no such massive parent galaxies 
are found in the Coma cluster
at z=0.02 is consistent with the down-sizing effect where the quenching
process of galaxies is shifted to lower mass systems as time goes on
, though still this small sample size cannot reject 
the possibility that the difference of the massive parent galaxies
in A851 and Coma is a simple statistical fluctuation.
More samples are required to draw a definitive conclusion about
EIGs; the evolution, the environmental effect, the variance among 
clusters and the effect of infalling groups.

\acknowledgments
We thank the anonymous referee for suggestions and comments.
We acknowledge Takeshi Urano for supporting our FOCAS observation.
We thank Yutaka Komiyama for suggestive comments.
This work has made use of the STARS2 archive,
SMOKA archive%
\footnote{http://smoka.nao.ac.jp/},
MAST archive, 
NED database%
\footnote{\url{http://nedwww.ipac.caltech.edu/}},
IRSA archive%
\footnote{\url{http://irsa.ipac.caltech.edu/data/SPITZER/Enhanced/Imaging/overview.html}},
and computer systems at ADC/NAOJ.

\onecolumn

\begin{figure}
\includegraphics[scale=0.6,bb=0 0 747 247]{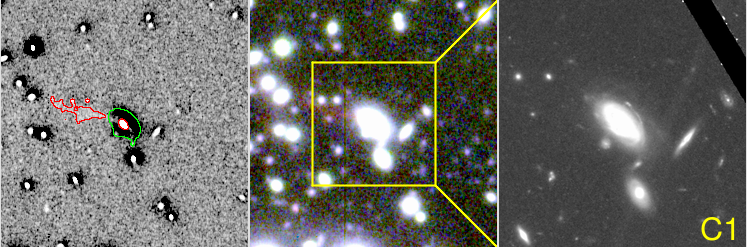}
\includegraphics[scale=0.6,bb=0 0 747 247]{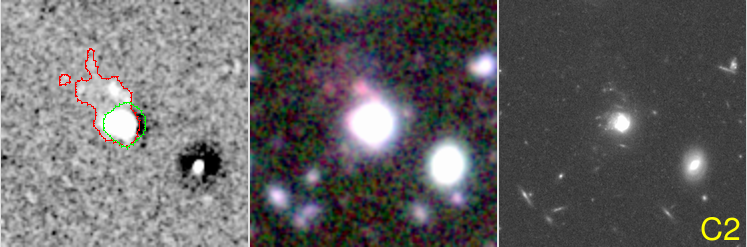}
\includegraphics[scale=0.6,bb=0 0 747 247]{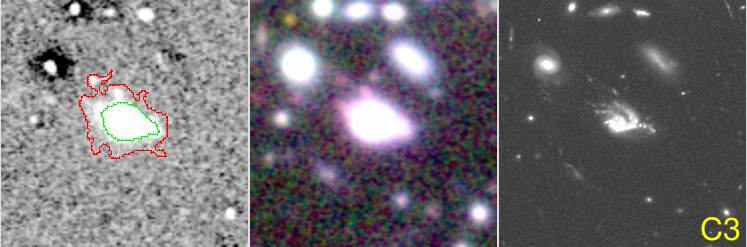}
\caption{
Cutouts of EIGs in A851.
North is up, east to the left.
(left) NB-z images in gray-scale. 
Image size of C1 is 50 arcsec ($\sim$ 270 kpc) square,
and others' are 25 arcsec ($\sim$ 140 kpc) square.
The gray-scale is the same for the nine EIGs.
The red contour shows the 3-$\sigma$ isophote 
extracted from the Gaussian-smoothed NB-z image.
Neighboring objects were omitted manually.
The green contour shows the 90-\% flux isophote of the parent galaxy
in z-band.
(middle) Three color composite image 
of R, z, and NB921 for blue, green and red. 
\Ha\ emitting regions are shown in red. 
The yellow square in C1 shows 25 arcsec square.
(right) An image of ACS in F816W.
Image size is 25 arcsec ($\sim$ 140 kpc) square.
}
\label{fig:postage}
\end{figure}

\addtocounter{figure}{-1}
\begin{figure}
\includegraphics[scale=0.6,bb=0 0 747 247]{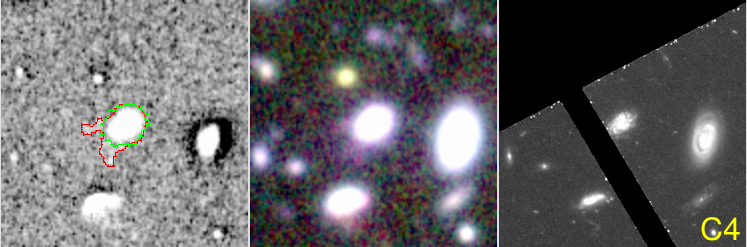}
\includegraphics[scale=0.6,bb=0 0 747 247]{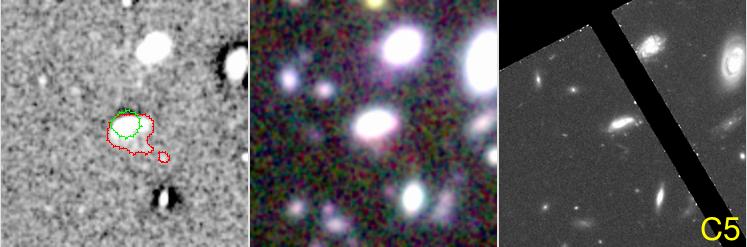}
\includegraphics[scale=0.6,bb=0 0 747 247]{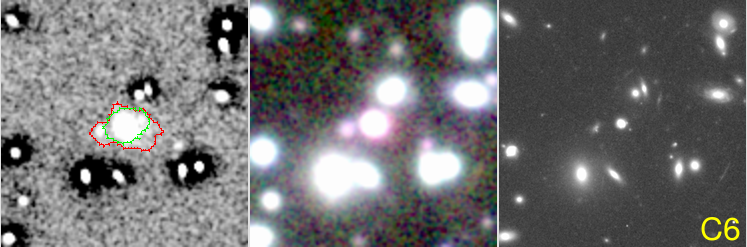}
\caption{Continued}
\end{figure}

\addtocounter{figure}{-1}
\begin{figure}
\includegraphics[scale=0.6,bb=0 0 747 247]{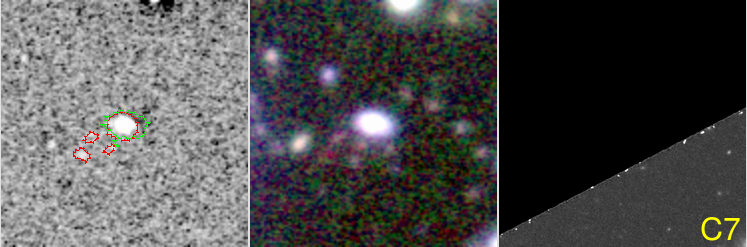}
\includegraphics[scale=0.6,bb=0 0 747 247]{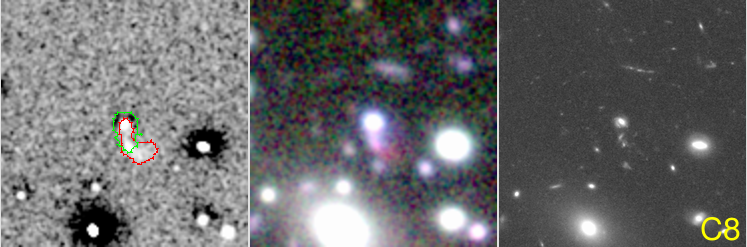}
\includegraphics[scale=0.6,bb=0 0 747 247]{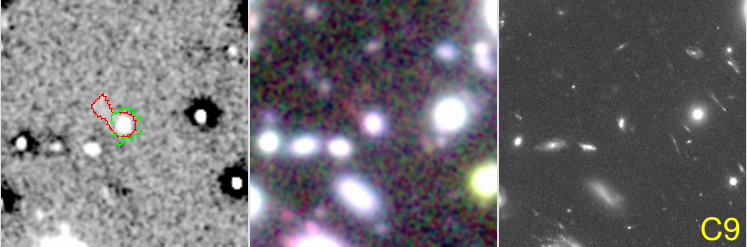}
\caption{Continued}
\end{figure}

\begin{figure}
\includegraphics[scale=0.6,bb=0 0 769 575]{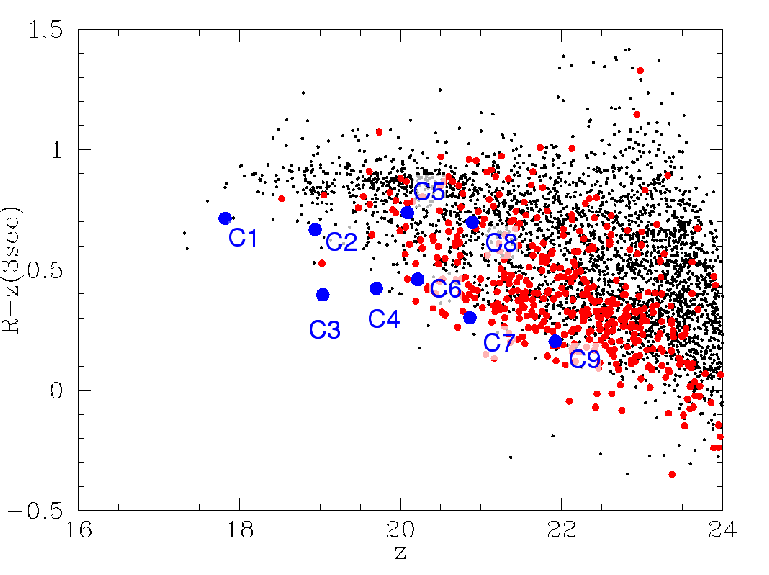}
\caption{Color magnitude diagram in A851 field.
The members selected from photometric redshift (phot-z) in \citep{Koyama2011} 
are plotted.
The filled red circles are z=0.4 emitters in \citet{Koyama2011}.
The parent galaxies of EIG are shown as filled blue circles.
}
\label{fig:CMD}
\end{figure}

\begin{figure}
\includegraphics[scale=0.6,bb=0 0 769 575]{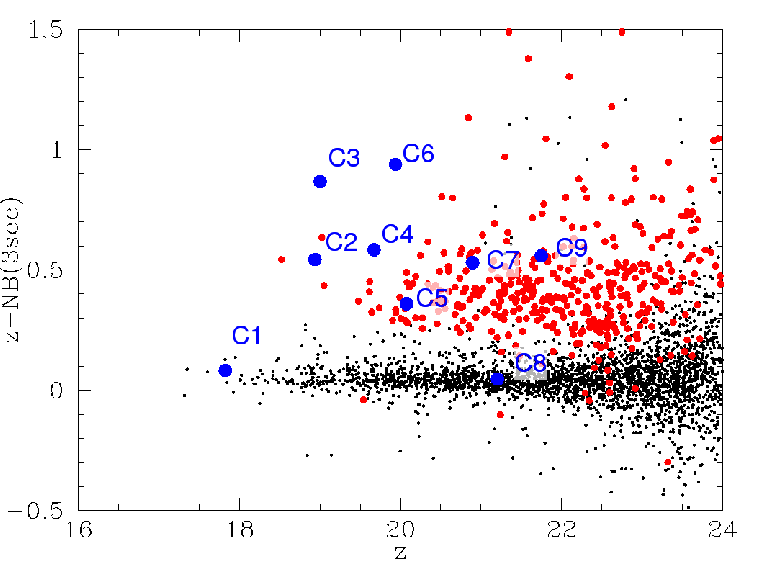}
\caption{Narrowband color excess of phot-z members 
of A851 in \citep{Koyama2011}.
The z-NB color is measured in revised data,
and is different from the \citet{Koyama2011} result.
The filled red circles are z=0.4 emitters in \citet{Koyama2011}.
The parent galaxies of EIG are shown as filled blue circles.
}
\label{fig:CMD_NB}
\end{figure}

\begin{figure}
\includegraphics[scale=0.6,bb=0 0 769 575]{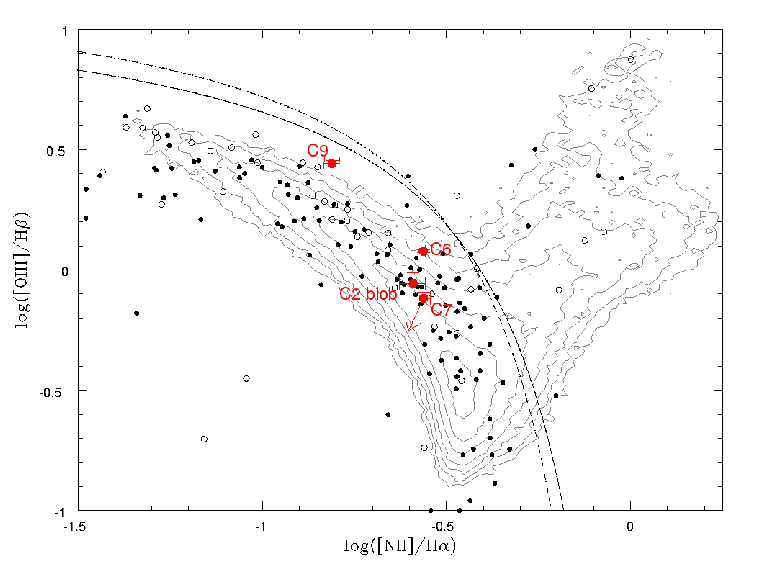}
\caption{Diagnostic line ratio plot of the bright spot of C2, C6, C7 and C9
(large, filled red circles with errorbars).
The solid and broken curves are the demarcation between \HII\ regions
and AGN by \citet{Kewley2013} at z=0.406 
and \citet{Kauffmann2003}, respectively.
Filled dots represent nearby field galaxies \citet{Jansen2000},
and open dots represent blue compact galaxies from \citet{Kong2002}.
The contour shows the distribution of SDSS galaxies
in the MPA-JHU DR7 release of spectrum measurements.
The data whose S/N of emission lines are larger than three are used,
and the number of galaxies are $4.0\times 10^5$.
The contour interval is a factor of $10^{1/4}$.
We also show the possible correction of 
background stellar absorption of C7 as a vector.}\label{fig:BPT}
\end{figure}

\begin{figure}
\includegraphics[scale=0.6,bb=0 0 768 574]{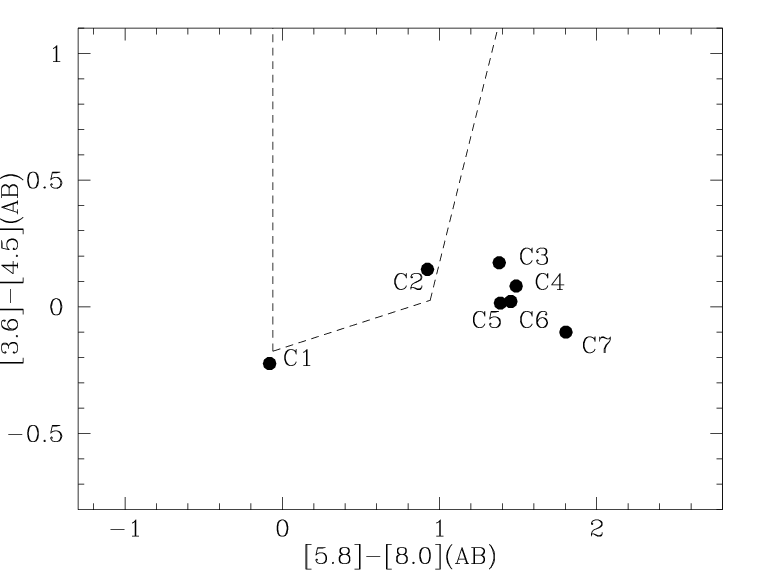}
\caption{IR two-color diagram as \citet{Stern2005}.
The broken line shows an empirical region of AGN.
Only C2 is in AGN region.}
\label{fig:IRcolmag}
\end{figure}

\begin{figure}
\includegraphics[scale=0.5,bb=0 0 889 889]{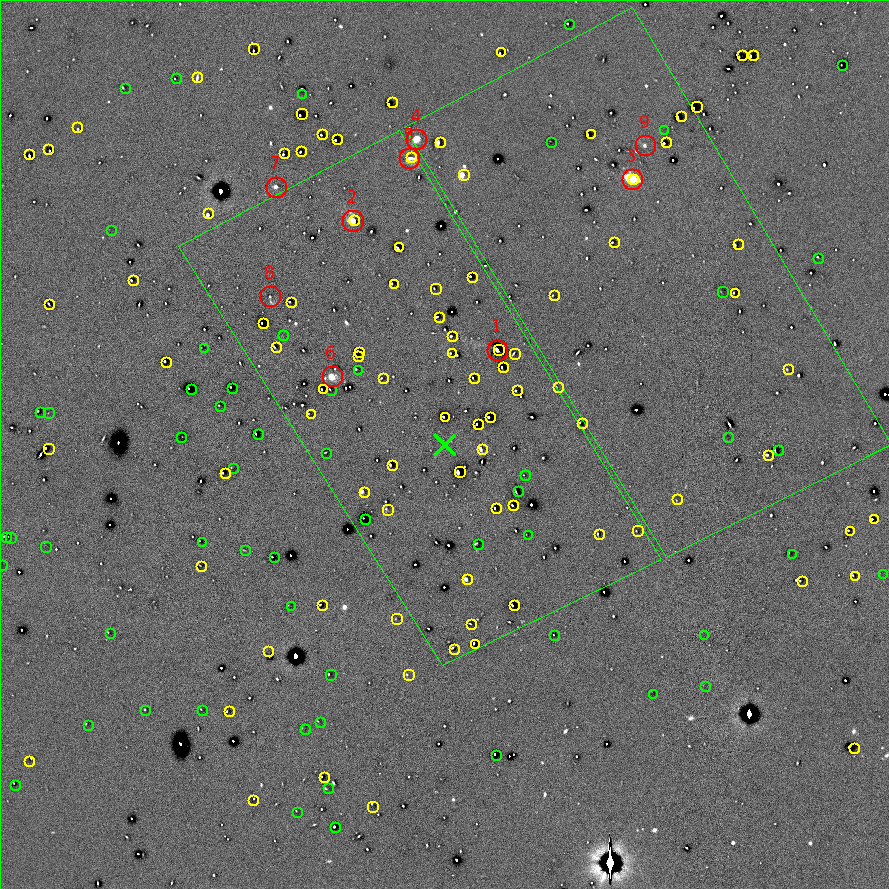}
\caption{Positions of the parents of EIGs.
The image size is 6 arcmin square, and the green X shows the cluster center.
The red circles with number show the parents.
The yellow and green circles are
the spectroscopic members and non-members of A851 in \citet{Oemler2009}.
The background image is NB-z. 
The field of view of the ACS data is shown as two tilted rectangles
in green.
}
\label{fig:map}
\end{figure}

\begin{figure}
\includegraphics[scale=0.6,bb=0 0 768 574]{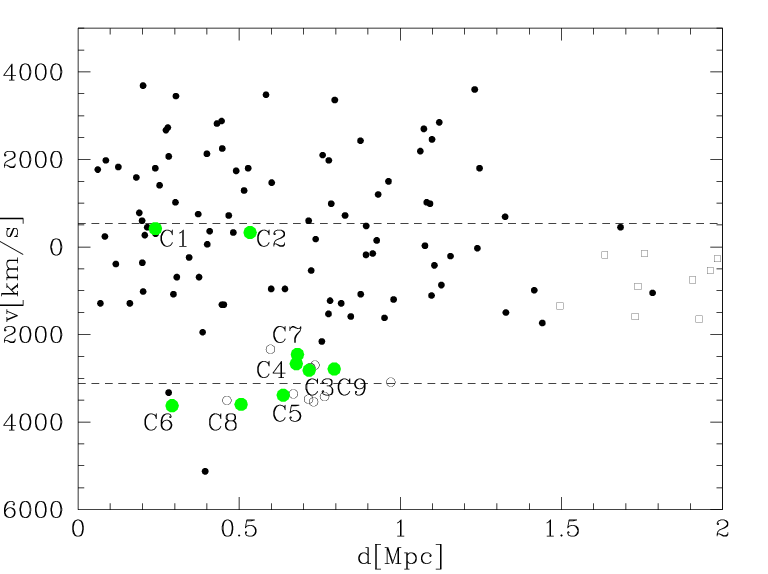}
\caption{Projected distance vs relative recession velocity 
to the cluster. 
The filled green circles with a label
are the EIG parents in this study.
The black dot, open circles, and open squares 
correspond to the Core, North and NW subcluster members 
by \citet{Oemler2009}, respectively.
The horizontal lines show the relative 
recession velocity of the Core and the North subclusters.
}
\label{fig:zd}
\end{figure}

\begin{figure}
\includegraphics[scale=0.4,bb=0 0 1004 1004]{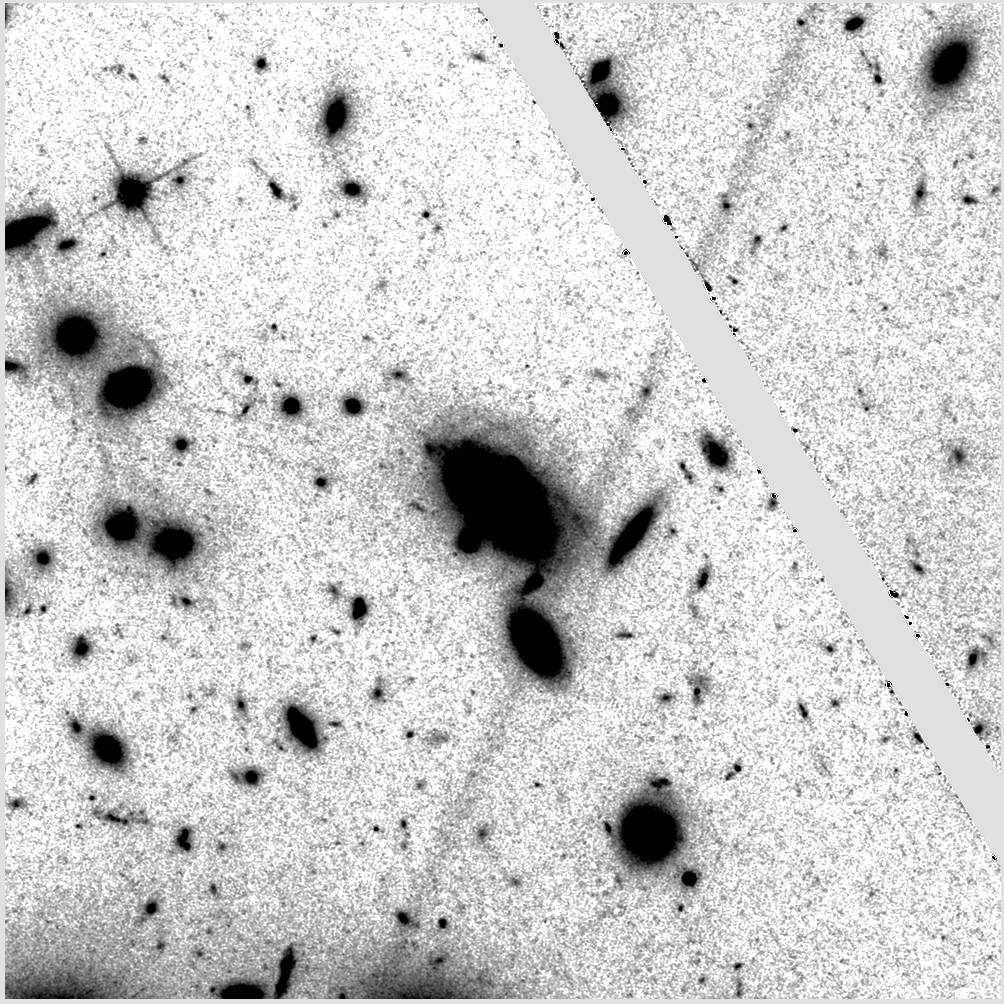}
\caption{
Cutouts of ACS F816W image around C1
with different contrast from Figure \ref{fig:postage} right
and inverted color.
Image size is 50 arcsec ($\sim$ 270 kpc) square.
North is up, east to the left.
A sign of stellar streams toward the northeast are
barely recognized.
}
\label{fig:C1}
\end{figure}

\begin{table}
\begin{tabular}{|c|c|c|c|c|c|c|l|l|l|}
\hline
ID & R.A.(2000) & Dec(2000) & m$_z$ & log & z & EIG & z-source\tablenotemark{a}& distor- & \Ha\\
   &            &           &      & $\left(\frac{{\rm M}_*}{{\rm M}_{\odot}}\right)$     &      & size & & tion& w/o \\
   &            &           &      &      &      &[kpc]   & && stars\\
\hline 
C1      &\RAform{09}{42}{55}{.91}&\Decform{+46}{59}{39.3}& 17.8 & 11.8 &0.4064& 86 &N-F8-22     &n&y\\
        &                        &                       &      &      &      &    &(ODK09-291) &&\\
C2      &\RAform{09}{43}{01}{.59}&\Decform{+47}{00}{31.4}& 18.9 & 11.2 &0.4061& 46 &ODK09-365   &y&y\\
        &                        &                       &      &      &      &    &            &&\\
C3      &\RAform{09}{42}{50}{.59}&\Decform{+47}{00}{47.9}& 19.0 & 10.5 &0.3956& 69 &N-F8-30     &y&n\\
        &                        &                       &      &      &      &    &(ODK09-354) &&\\
C4      &\RAform{09}{42}{59}{.08}&\Decform{+47}{01}{04.3}& 19.7 & 10.5 &0.3961& 27 &N-F8-26     &y&n\\
C5      &\RAform{09}{42}{59}{.37}&\Decform{+47}{00}{56.5}& 20.1 & 10.8 &0.3937& 20 &ODK09-367   &y&n?\\
C6      &\RAform{09}{43}{02}{.46}&\Decform{+46}{59}{28.6}& 20.2 & 10.2 &0.3929& 31 &this study  &n&y\\ 
C7      &\RAform{09}{43}{04}{.63}&\Decform{+47}{00}{45.3}& 20.9 & 10.0 &0.3968& 32 &this study  &?&y\\
        &                        &                       &      &      &      &    &            &&\\
C8      &\RAform{09}{43}{04}{.88}&\Decform{+47}{00}{01.0}& 20.9 & 10.6 &0.393   & 24 &this study&n?&y\\
C9      &\RAform{09}{42}{50}{.05}&\Decform{+47}{01}{01.4}& 21.9 & 9.4  &0.3957& 21 &this study  &n&y\\
\hline
\end{tabular}
\caption{EIG parents in A851}
\label{tab:target}
\tablenotetext{a}{Redshift reference and ID; N:\citet{Nakata2014},
ODK09:\citet{Oemler2009}
}
\end{table}

\begin{table}
\begin{tabular}{|c|c|c|}
\hline
name & \NII/\Ha & \OIII/\Hb \\
\hline
C2 bright spot & 0.25$\pm$0.02 &0.88$\pm$0.09\\
C6 & 0.274$\pm$0.004 &1.20$\pm$0.02\\
C7 & $<$ 0.27$\pm$0.01 & $<$ 0.76$\pm$0.04\\
C9 & 0.15$\pm$0.01 &2.77$\pm$0.08\\
\hline
\end{tabular}
\caption{Diagnostic line ratios from FOCAS spectra}
\label{tab:lineratio}
\end{table}

\begin{table}
\begin{tabular}{|c|c|c|c|c|c|}
\hline
object & B-R\tablenotemark{a} & R-z\tablenotemark{a} & [3.6]-[4.5]\tablenotemark{b} & [5.6]-[8.0]\tablenotemark{b} & 24$\mu$m[mJy]\tablenotemark{c} \\
\hline 
C1     &0.71 &1.90 &  -0.22 &-0.08 & 0.30 \\
C2     &0.67 &1.41 &   0.15 & 0.92 & 3.48 \\
C3     &0.40 &0.81 &   0.17 & 1.38 & 0.90 \\
C4     &0.42 &0.98 &   0.08 & 1.49 & ...  \\
C5     &0.74 &1.52 &   0.02 & 1.38 & 0.56 \\
C6     &0.46 &0.80 &   0.02 & 1.45 & 1.09 \\
C7     &0.30 &1.06 &  -0.10 & 1.80 & ...  \\
C8     &0.70 &1.78 &  -0.31 & ...  & ...  \\
C9     &0.20 &0.93 &  -0.27 & ...  & ...  \\
\hline
\end{tabular}
\caption{Color and flux of EIG parents}
\label{tab:color}
\tablenotetext{a}{Color of 3 arcsec aperture magnitude}
\tablenotetext{b}{Color of 5.8 arcsec aperture magnitude}
\tablenotetext{c}{aperture flux density}
\end{table}

\clearpage

\appendix
\section{Fringe correction of Suprime-Cam data}

The z, NB921, and NB912 data of Suprime-Cam suffer 
fringe pattern of atmospheric emissions.
The correction was performed as follows.

First, the night sky images are flat-fielded using dome flats.
As the fringe pattern is approximately proportional 
to the sky brightness,
we normalize the flat-fielded skys by the sky count.
Then, a ``median normalized fringe'' is constructed 
in a two-pass procedure.
In the first pass, median image of the normalized flat-fielded sky 
is constructed.
It is affected by objects in the night sky images.
Such objects are detected by comparing each normalized image
with the median image.
In the second pass, the detected objects are masked in 
normalized sky images, and their median is constructed.
We use the output as the ``median normalized fringe''.

\begin{figure}[htb]
\includegraphics[scale=0.33,bb=0 0 1533 471]{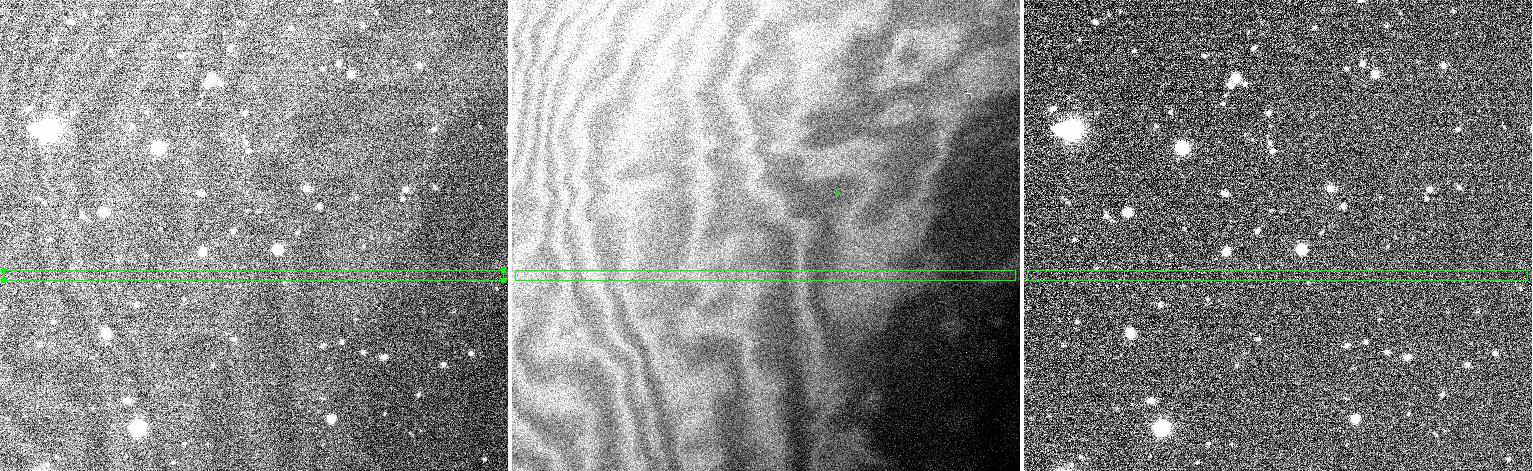}
\caption{
Example of the automatic fringe subtraction.
A cutout of 1000x1000 pixels is shown.
(left) flat-fielded image 
(middle) median normalized fringe,
(right) fringe corrected image.
The green rectangle shows the region where the 
profile shown in \ref{fig:fringe2} is measured.
}
\label{fig:fringe1}
\end{figure}

\begin{figure}[htb]
\includegraphics[scale=0.33,bb=0 0 768 574]{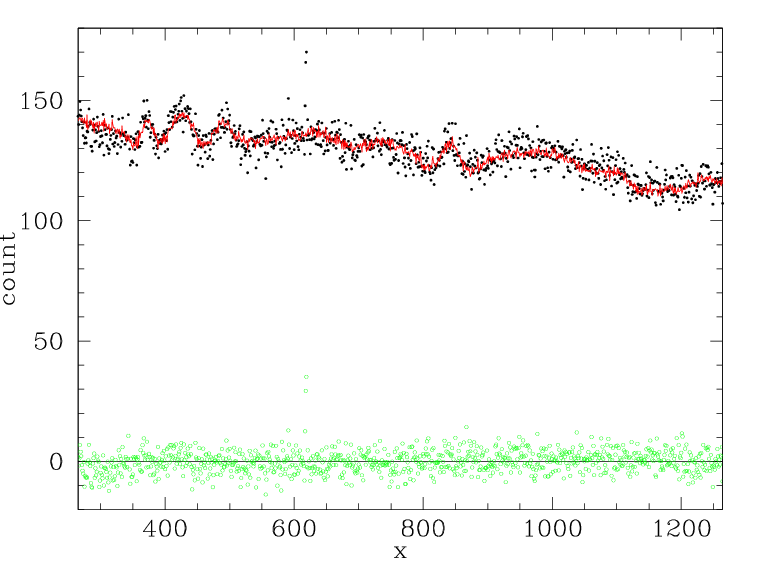}
\caption{
The spatial profile of the pixel value in Figure \ref{fig:fringe1}.
At each x, y-is binned in 20 pixels.
Black, red, and green correspond to
the mean pixel value in flat-fielded image($s(x,y)$),
estimated model ($a f(x,y)+b$), and the residual, respectively.
}
\label{fig:fringe2}
\end{figure}

In the real data, the fringe pattern is not 
always proportional to the background sky brightness.
The scaling of the median normalized fringe is 
therefore calculated for each sky image of the target.
We used two robustization techniques for the fitting
\[
s(x,y)=a f(x,y)+b,
\]
where $s(x,y)$ is the target sky image, $f(x,y)$ is 
the median normalized fringe, 
and the coefficients $a$ and $b$ are to be solved.
First, $f(x,y)$ is sorted and binned into $N$ (we adopted $10^4$) 
data points and the median is taken $f'(n)$.
The median of the corresponding $s(x,y)$ is assigned as $s'(n)$.
Note that the data points are $\sim 8\times 10^6$ in an image,
and $\sim 800$ points are obtained.
Then, 
\[
s'(n)=a f'(n)+b,
\]
is fitted by the least median of squares method
\citep{Rousseeuw1984}.
The example of the automatic fringe subtraction is shown as Figures \ref{fig:fringe1} and \ref{fig:fringe2}.
The higher order residual remains, but is negligible 
in this study. 

\clearpage

\section{The effect of non-uniformity of narrowband filters}

As the prime focus of the Subaru telescope has small F-ratio
(F/2), the response function of narrow-band filters 
is bell-shaped in general.
The detectability of \Ha\ excess from narrow-band imaging 
is therefore highly dependent on the redshift of the \Ha\ emission.

Meanwhile, a spatial non-uniformity of the transmission of NB921 filter 
was reported by \citet{Hayashino2003}.
\citet{Hayashino2003} also measured the transmission of NB912 filter and 
presented this on their web page (not published).
We examined the possibility that the non-detection of EIGs in
CL0024+17 and outer part of A851 is affected by this non-uniformity.

\begin{figure}[hb]
\includegraphics[scale=0.4,bb=0 0 768 574]{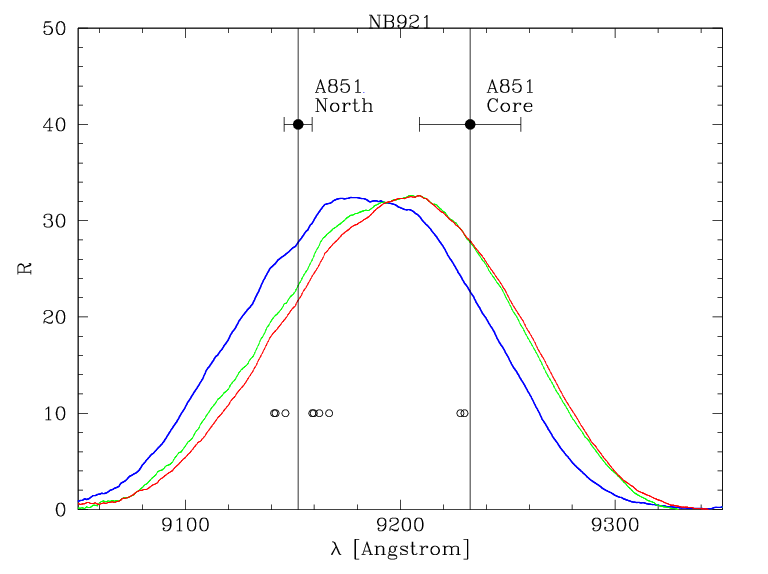}
\caption{
Total transmission curves of NB921 filter and optics and atmospheric 
extinction. Blue, green, and red lines represent 
0, 11.5 and 22.8 arcmin from the center of the filter, respectively
\citep{Hayashino2003}.
The wavelengths of \Ha\ at redshift of two subclusters in A851 
are shown as vertical lines. The horizontal error bar shows
1$\sigma$ of the velocity dispersion of the subcluster.
Open circles show \Ha\ wavelength of EIG parents.}
\label{fig:A851trans}
\end{figure}

In Figure \ref{fig:A851trans},
the total throughput at three different positions
and the \Ha\ wavelength at A851 subclusters are shown.
The total throughput was calculated by 
multiplying the filter transmission 
by the quantum efficiency(QE) of CCDs,
the transmittance of the primary focus corrector,
the reflectivity of Primary mirror,
and the extinction of a model atmosphere at airmass=1
\citep{Yagi2013}.
The total throughput of NB921 at the center of the filter 
has an offset by about 15\AA\ bluer than that at 11.5 arcmin 
from the center.
The difference of the transmission may cause $\lesssim 40\%$ 
variation of \Ha\ throughput at different position in the field,
depending on the redshift.

In the Figure, the \Ha\ wavelength of EIG parents of A851 were 
shown as open circles.
Since the EIG parents were found within 3.3 arcmin from the 
filter center, the total throughput at the filter center 
(blue in Figure \ref{fig:A851trans}) would be affected.
The redshift variation of the EIG parents 
would therefore make the variation of the total throughput by 35\% p-p;
EIGs in the North subcluster have better \Ha\ throughput 
than those in the Core subcluster around the filter center.

In the outer part of A851, we can expect that 
the possible undetected EIG parents would have redshifts 
comparable to the Core subcluster, since the mass of the subcluster and 
the number of the members are larger than those of the North subcluster.
Then, the throughput of \Ha\ should be higher at the outer part of 
the filter, as shown in Figure \ref{fig:A851trans}.
Moreover, the throughput of \Ha\ at the redshift of the Core subcluster
at the outer part is comparable to that at the redshift of the North
subcluster around the center.
We can therefore conclude that it was not due to the
non-uniformity of NB921 that the detection of EIG parents was
only around the center.

\begin{figure}[hb]
\includegraphics[scale=0.4,bb=0 0 768 574]{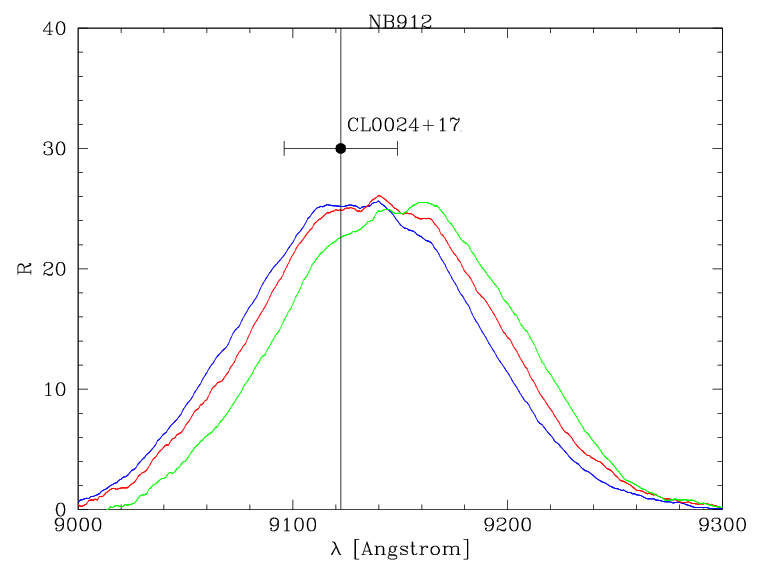}
\caption{
Same as figure \ref{fig:A851trans} but of NB912 and CL0024+17.
}
\label{fig:CLtrans}
\end{figure}

NB912 also shows such offset by $\sim$ 10\AA\ at 11.5 arcmin
as shown in Figure \ref{fig:CLtrans}.
However, as the wavelength of \Ha\ at the redshift of CL0024+17 
is near the peak of the transmission, 
the offset does not cause a large difference of the throughput of 
\Ha\ between at the center and at 11.5 arcmin from the center
compared with the effect of the variation of the redshift.

\end{document}